\documentclass[twocolumn]{aastex631}
\usepackage[utf8]{inputenc}

\def\cycleday{${\rm d}^{-1}$}

\begin{document}

\title{A Surprising Periodicity Detected During a Super-outburst of V844~Herculis by TESS}

\author[0009-0008-6389-9398]{Anousha Greiveldinger}
\affiliation{Department of Physics and Astronomy, University of Notre Dame, Notre Dame, IN 46556, USA}

\author[0000-0003-4069-2817]{Peter Garnavich}
\affiliation{Department of Physics and Astronomy, University of Notre Dame, Notre Dame, IN 46556, USA}

\author[0000-0001-7746-5795]{Colin Littlefield}
\affiliation{Department of Physics and Astronomy, University of Notre Dame, Notre Dame, IN 46556, USA}
\affiliation{Bay Area Environmental Research Institute, Moffett Field, CA 94035 USA}


\author[0000-0001-6894-6044]{Mark R. Kennedy}
\affiliation{School of Physics, University College Cork, Cork, Ireland}
\affiliation{Jodrell Bank Centre for Astrophysics, Department of Physics and Astronomy, The University of Manchester, Manchester M13 9PL, United Kingom}

\author[0000-0003-4814-2377]{Jules P. Halpern}
\affiliation{Department of Astronomy, Columbia University, New York, NY 10027  USA}

\author[0000-0002-4964-4144]{John R. Thorstensen}
\affiliation{Department of Physics and Astronomy, Dartmouth College, Hanover, NH 03755 USA}

\author[0000-0003-4373-7777]{Paula Szkody}
\affiliation{Department of Astronomy, University of Washington, Seattle, WA 98195, USA}

\author[0000-0001-6249-6062]{Arto Oksanen}
\affiliation{American Association of Variable Star Observers, 49 Bay State Rd, Cambridge, MA 02138, USA}

\author[0000-0003-1085-4416]{Rebecca S. Boyle}
\affiliation{Department of Physics and Astronomy, University of Notre Dame, Notre Dame, IN 46556, USA}

\begin{abstract}

 We identify a previously undetected periodicity at a frequency of 49.08$\pm 0.01$~\cycleday\ (period of 29.34$\pm0.01$ minutes) during a super-outburst of V844~Her observed by TESS.
 V844~Her is an SU~UMa type cataclysmic variable with an orbital period of 78.69 minutes, near the period minimum. The frequency of this new signal is constant in contrast to the superhump oscillations commonly seen in SU~UMa outbursts.  We searched without success for oscillations during quiescence using MDM, TESS, and XMM-Newton data. The lack of a periodic signal in the XMM light curve and the relatively low X-ray luminosity of V844~Her suggests that it is not a typical IP. We consider the possibility that the 29~min signal is the result of super-Nyquist sampling of a Dwarf Nova Oscillation with a period near the 2-minute cadence of the TESS data. Our analysis of archival AAVSO photometry from a 2006 super-outburst supports the existence of a 29~min oscillation, although a published study of an earlier superoutburst did not detect the signal. We compare the X-ray properties of V844~Her with short orbital period intermediate polars (IP), V1025~Cen and DW~Cnc. We conclude that the new signal is a real photometric oscillation coming from the V844~Her system and that it is unlikely to be an aliased high-frequency oscillation. The steady frequency of the new signal suggests that its origin is related to an asynchronously rotating white dwarf in V844~Her, although the precise mechanism producing the flux variations remains unclear. 
 
    
\end{abstract}


\section{Introduction}

Cataclysmic variables (CV) are binary star systems consisting of an accreting white dwarf (WD) and a late-type secondary star. In CVs, the secondary star generally fills its Roche lobe and transfers mass to the WD through a stream. An intermediate polar (IP) is a subclass of CV in which the primary is midly magnetic. The accretion process results in a wide range of periodic and quasi-periodic oscillations on time scales of the orbital period and the WD spin period. In high mass-ratio systems, for example, a dynamical accretion disk instability results in ``superhump'' oscillations seen at frequencies within a few percent of the system's orbital period \citep{whitehurst88,osaki89}. Accretion onto a rapidly spinning, magnetic WD, as seen in IPs, may result in the optical and X-ray modulations at the WD spin period or the beat period between the spin and orbital frequencies \citep{patterson94}.  

CVs can also show rapid photometric oscillations with periods ranging from tens to hundreds of seconds. This variability is called Dwarf Nova Oscillations (DNO) because they are often detected when an accretion disk has transitioned to a bright, hot, highly viscous state seen as a dwarf nova outburst. The complex observational characteristics of DNO were reviewed by \citet{warner04} and \citet{warner08}, and include long-period DNO (lpDNO) and associated quasi-periodic oscillations (QPO). These oscillations are thought to result from an interaction between the accretion disk and a spinning, magnetic WD.  DNO and related oscillations seen in CVs appear to have corresponding phenomena in neutron star and black hole binaries \citep{warner03}.

The 1800~s and 120~s cadences of TESS observations are ill-suited for the study of DNO phenomena that typically show oscillation frequencies of 10~s to a few 100~s. However, the extremely precise cadence of TESS leads to Nyquist aliasing where an oscillation at a frequency greater than the Nyquist frequency appears as a lower frequency signal. This ``Super-Nyquist'' technique has been applied to asteroseismology with Kepler \citep{murphy15}, and it may be useful in identifying DNO-like oscillations with TESS.

V844~Her has been classified as an SU~UMa type dwarf novae (DN) due to evidence of longer and brighter super-outbursts, as well as normal outbursts. It has a very short orbital period of 78.69~min \citep{thorstensen02}. \citet{oizumi07} noted that normal outbursts from V844~Her are rare and that the time between super-outbursts is about 300~days. 

\begin{figure}
    \centering
    \includegraphics[width=\columnwidth]{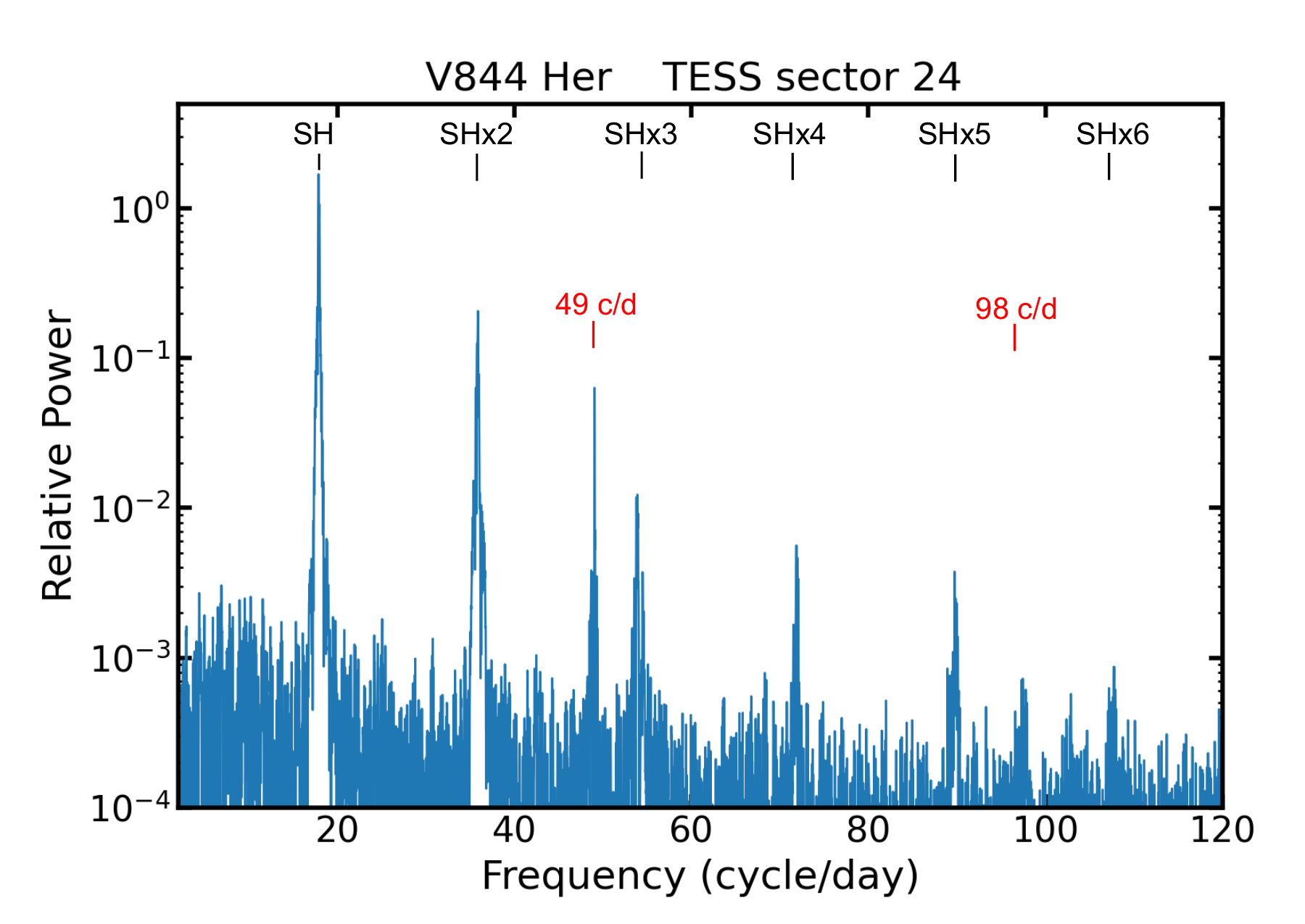}
    \caption{The Lomb-Scargle periodogram for V844~Her calculated from the sector~24 TESS light curve. The strong superhump signal is seen at 18~\cycleday (80~min), and the its harmonics are detected out to 108~\cycleday (13.3~min). The strong peak at 49~\cycleday\ (29.34~min) has not been identified. A weak signal near 98~\cycleday\ (14.7~min) may be a harmonic of the unidentified oscillation.   }
    \label{pow}
\end{figure}

\section{Data}

\subsection{TESS} 

V844 Her, TIC~39950064, was observed by TESS in the fast cadence mode (120~s exposures) during sectors 24 and 25. Observations were recorded in sector 24 from April 16 to May 12, 2020, and V884 Her was visible on Camera~1, CCD~1. The star was in super-outburst over sector~24. In sector 25, observations were conducted from May 14 to June 8, 2020, and V884~Her was visible on Camera~1, CCD~2. V844~Her was noted by \citet{kato22} to be in super-outburst during sector~24. \citet{kato22} analyzed the behavior of superhumps in the data, but we focus on the periodicity occuring at 49 \cycleday\ that was not included in the analysis. V844~Her was again observed over sectors 51 and 52 covering the dates April 22 through June 13, 2022. The star was in quiescence over these dates. 

The data were acquired using the Lightkurve python package \citep{lightkurve18} and periodograms were calculated using the Lomb-Scargle technique \citep[L-S hereafter;][]{Lomb76, Scargle82}. The periodogram of sector~24 is displayed in Figure~\ref{pow}. It clearly shows the superhump (SH) oscillations at 80.45$\pm 0.45$~min (17.9$\pm 0.1$~\cycleday) and their harmonics. The SH signals are broadened by the classic SU~UMa frequency variations as described by \citet{kato22} for this same data set. Any orbital modulation would appear at 78.69~min (18.30~\cycleday), but it is not easily seen in the TESS data.

A fairly strong signal is obvious at 49.08$\pm 0.01$~\cycleday, or 29.34 $\pm 0.01$~minutes, (29~min hereafter), but it appears unrelated to the SH oscillations or orbital period. An excess of power is seen at 98~\cycleday\ (14.7~min) which may be a weak harmonic of this unidentified signal. The peak at 29~min in the periodogram is narrow and unresolved, suggesting a nearly constant frequency during sector~24.

\begin{figure*}
    \centering
    \includegraphics[scale=0.7]{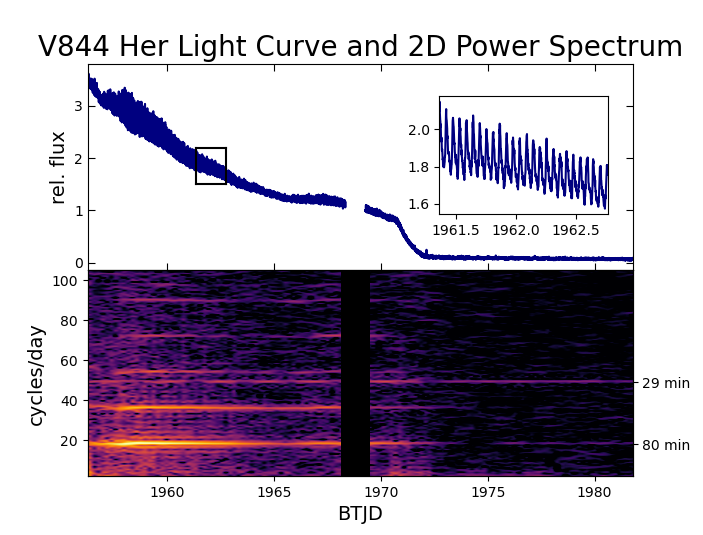}
    \caption{{\bf Top:} A portion of the TESS light curve of V844~Her during sector~24. The super-outburst is seen fading from a maximum at the start of the observations. The development of classical superhump oscillations \citep{kato22} is seen. {\bf Bottom:} The time-resolved power spectrum is displayed over sector~24. The amplitude of the superhump oscillations decays with the fading of the super-outburst. The strength of the 29~min signal fades over the sector, but more slowly than that of the superhumps. By the end of the sector, the 29~min oscillation is the strongest signal seen. The new signal and the primary superhump oscillation are marked on the right side of the figure with their periods of 29~min and 80~min, respectively.}
    \label{lc}
\end{figure*}

The variability of the oscillations is tested using a time-resolved power spectrum shown in Figure~\ref{lc}. The Lomb-Scargle periodogram is calculated in a sliding 1-day window. The amplitudes of the SH oscillations and harmonics are seen to decline along with the fading of the super-outburst. The amplitude of the 29~min signal fades more slowly than the SH oscillations and remains detectable to the end of sector~24 and into sector~25.

\subsection{MDM Photometry}

V844~Her was observed with the MDM 1.3~m McGraw-Hill telescope on four consecutive nights starting on May~28, 2022 (UT). A $GG420$ long-pass filter was used with the MDM Templeton detector. To minimize overheads, only a 130-arcsec section of the sky was recorded for analysis. The star was observed for three hours each night with a 20~s exposure time resulting in an average cadence of 23~s. 

V844~Her had an approximate $V$-band magnitude of 17.9 while observed at MDM. This is consistent with its typical quiescent brightness. The light curves revealed a flickering with an amplitude of $\pm 0.15$~mag over a timescale of tens of minutes.  

\subsection{AAVSO}

Searching the American Association of Variable Star Observers (AAVSO) International Database \footnote{https://www.aavso.org/LCGv2/}, we identified a long photometric time-series of V844~Her obtained during a super-outburst. The unfiltered CCD photometry was taken over three consecutive nights beginning April~29, 2006. The time-series had a typical cadence of 30~s between exposures, and a total of 771 measurements were made over the three nights. No error estimates were included in the data table, but the scatter of the points near minima of the superhumps implies a photometric precision of approximately 0.02~mag.

Observations began five nights after the start of the super-outburst when V844~Her was at 12.9~mag. The superhump amplitude was initially 0.2~mag. A Lomb-Scargle periodogram of the photometry is strongly impacted by the sampling window, but the fundamental superhump signal and its first harmonic are clearly detectable. Further analysis of this data is discussed in Section~3.

\subsection{XMM}

V844~Her was serendipitously observed by \emph{XMM-Newton} as part of observation ID 0784521701, which was intended to target the galaxy cluster Abell~2199. It was observed on March 22, 2017, at 01:59:14 (UT) for 31.9 ks. Both of the EPIC-MOS sensors \cite{turner01} and the EPIC-pn sensor \cite{struder01} were operated in Full Frame Mode with a Medium Filter. The two Reflection Grating Spectrometer (RGS) instruments \cite{Rasmussen00} were operated in  Spectroscopy HER + SES mode without a filter. The Optical Monitor (OM) \cite{Cordova00} was operated in Image Mode and used three different filters throughout the observation: U, UVW1, and UVM2.

For comparison with known and suspected IPs, we re-analyzed archival XMM observations of V1025~Cen, DW~Cnc and, WZ~Sge.

V1025~Cen, observation ID 0673140501, was executed by \emph{XMM-Newton} on Jan 1, 2012, at 22:21:43 (UT) for a total of 16.5~ks. The EPIC-MOS sensors \citep{turner01} were operated in Small Window Mode with a Medium filter and the EPIC-pn sensor \citep{struder01} was operated in Large Window Mode with a Medium filter. The OM \citep{Cordova00} was operated in the Image + Fast mode with the UVM2 filter and both the RGS instruments \citep{Rasmussen00} were operated in the Spectroscopy HER + SES mode without a filter.

DW~Cnc was observed by \emph{XMM-Newton} on April 2, 2012, at 09:06:03 (UT) for a total of 9.4~ks with observation ID 0673140101. The three EPIC sensors and two RGS sensors were operated in the same modes as for V1025 Cen. The OM was operated in Fast mode with the UVM2 filter. 

WZ Sge was observed by \emph{XMM-Newton} on May 16, 2003, at 14:52:07 (UT) for a total of 9.9~ks with observation ID 0150100101. The EPIC-MOS sensors were operated in Large Window Mode with a Thin1 filter and the Epic-pn sensor was operated in Full Frame mode with the same filter. The OM was in Image + Fast mode with the UVW1 filter and the RGS instruments were both run in the Spectroscopy HER + SES mode without a filter.

The data were collected from the \emph{XMM-Newton} Science Archive using RISA analysis of EPIC exposures. The bin size was set to the default value of 100 seconds and counts were calculated by averaging the number of counts per bin. The data were reduced and analyzed using \emph{XMM}-Science Analysis System (SAS v19.0.0). The observation raw data files (ODFs) were downloaded from the \emph{XMM-Newton} Science Archive and the current calibration files (CCF) were retrieved using the \emph{cifbuild} command. The redistribution matrix files (RMF) were generated using the \emph{rmfgen} and \emph{rmfset} commands. Then, the spectrum, background, RMF, and ancillary response files (ARF) files were grouped for easier handling in \emph{XSPEC} \citep{arnaud96}.

\section{Analysis}

\subsection{Optical Variations}

The 29~min signal observed in the TESS observations has not been reported before. An extensive photometric study of V844~Her during a 1997 outburst by \citet{thorstensen02} did not show any excess signal around 29~min (their Figure~6). We note that the SH harmonic at 26.7~min, near the new periodicity, was also not seen, but harmonics at 40 and 20~min were well detected. From the AAVSO archive, the outburst studied by \citet{thorstensen02} reached a maximum visual magnitude of 12.2, with the bright state lasting for 15~days before fading to quiescence over 1-2 days. The brightness and length are consistent with it being a superoutburst.

With no ground-based confirmation, we must consider that the TESS signal could be an artifact or a periodicity from a source near V844~Her.

Its detection in TESS but not in ground data also suggests that it could be an oscillation with a period shorter than the TESS cadence and appearing as a super-Nyquist sampled signal at the difference between its true frequency and the TESS sampling frequency. 

Alternatively, TESS is highly sensitive to weak, but long-lasting signals, suggesting that ground-based observations may have not been sufficient to detect the signal in the past. Thus, when possible, we searched for variations at a 29~min period in several other datasets, as well as for signals at the super-Nyquist frequencies around 2~minutes (720~\cycleday).

\begin{figure}
    \centering
    \includegraphics[width=\columnwidth]{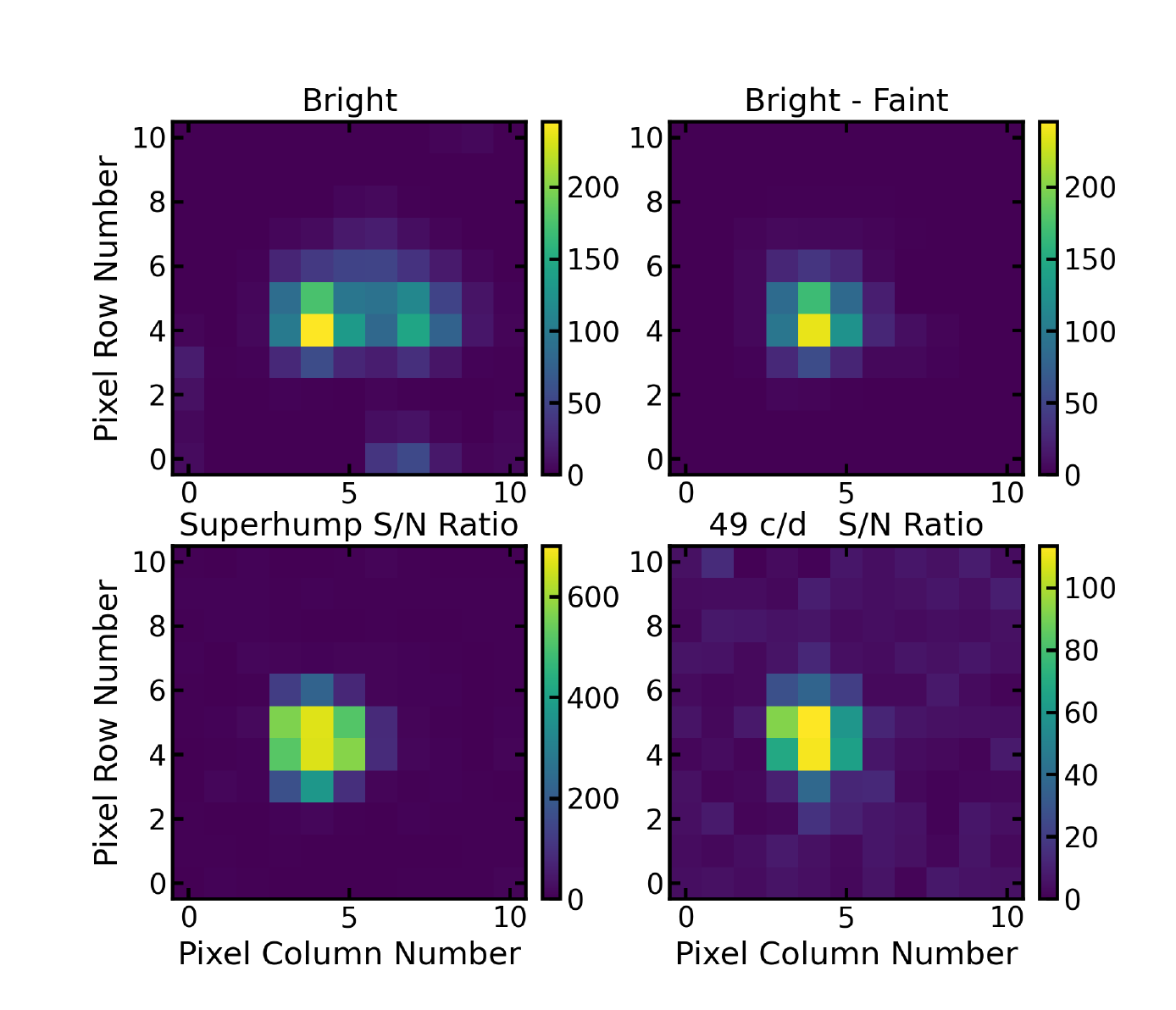}
    \caption{An analysis of the Target Pixel Files (TPF) showing the location V844~Her and the location of the oscillations. {\bf Top-Left:} The average TPF over the first 9~days of the outburst.  {\bf Top-Right:} The difference between the brightest 9~days and the average of the final 10~days of sector 24. As the outburst had faded significantly, the residuals isolate the light distribution of V844~Her during sector~24. {\bf Lower-Left:} The strength of the superhump signal as measured by the S/N ratio of the periodogram peak. As expected, the strongest signal coincides with the light distribution of V844~Her. {\bf Lower-Right:} Shows the strength of the 29~min signal in the TPF images. The position of highest S/N ratio of the 29~min oscillation matches the location of V844~Her and the superhump signal strength. } 
    \label{bckgrndStar}
\end{figure}

\subsubsection{Localization of the Oscillation Signals}

The aperture used to extract the light curve of V844~Her is likely to include light from one or more nearby stars. The 29~min signal may be originating from one of these contaminating stars \citep[e.g.][]{higgins23}. The average target pixel file (TPF) image while V844~Her was in outburst is shown in the top-left panel of Figure~\ref{bckgrndStar}. Several sources are visible in the TPF.

To determine the precise location of the light associated with V844~Her, we averaged the TPF images from the last ten days of sector~24 when the outburst had significantly faded. We then subtracted the faint epochs from the outburst epochs, and this result is shown in the top-right panel of Figure~\ref{bckgrndStar}. Stars that have remained constant over the sector are removed, while the positive residuals show the flux distribution coming exclusively from V844~Her. 

The superhump oscillations are undoubtedly coming from V844~Her, and we localize the oscillations by calculating the L-S periodogram for each TPF pixel and estimating the signal-to-noise ratio (S/N) of the first superhump harmonic at 40~min. The noise level is estimated by taking the median value of the power on each side of the signal peak. The signal is assumed to be the maximum of the periodogram between $35.5< f < 36.3$, where $f$ is the frequency in \cycleday . The strength of the superhump signal in each pixel is shown in the lower-left panel of Figure~\ref{bckgrndStar}. As expected, the highest S/N of the superhump power corresponds to the position of V844~Her. 

To determine the location of the 29~min signal, we repeat the calculation done previously, but with the peak power estimated from the range $48.7< f < 49.8$~\cycleday , and the noise is estimated from the median of frequencies on each side of the peak. The distribution of the S/N from the 29~min signal is shown in the lower-right panel of Figure~\ref{bckgrndStar}. The centroid of the 29~min signal is consistent with the superhump signal and the position of V844~Her. 

\subsubsection{Behavior of Other Stars on the TESS Detectors}

The periodic thruster firing of the Kepler spacecraft during the extended K2 mission induced periodic signals into its stellar photometry \citep[e.g.][]{clarke21}. TESS has been very stable in comparison, but there may be spurious signals added to the data through electronic noise, and this may depend on the detector or be time-dependent. To test if the 29~min signal is a detector artifact, we selected several stars near V844~Her that were recorded on Camera~1/CCD~1 during sector~24 using the 2-minute cadence. 

We calculated L-S periodograms for two dozen stars that met the above criteria. The stars brighter than T$_{mag}\approx 10 $ consistently showed excess power spread between 131 and 144~min (between 10 and 11~\cycleday). Time-resolved power spectra revealed that this excess was concentrated over the dates 1964$<BTJD<1968$ and this time range also corresponds to a general increase in photometric noise. No excess power detected in nearby stars was seen around periods of 29~min.

\begin{figure}
    \centering
    \includegraphics[width=\columnwidth]{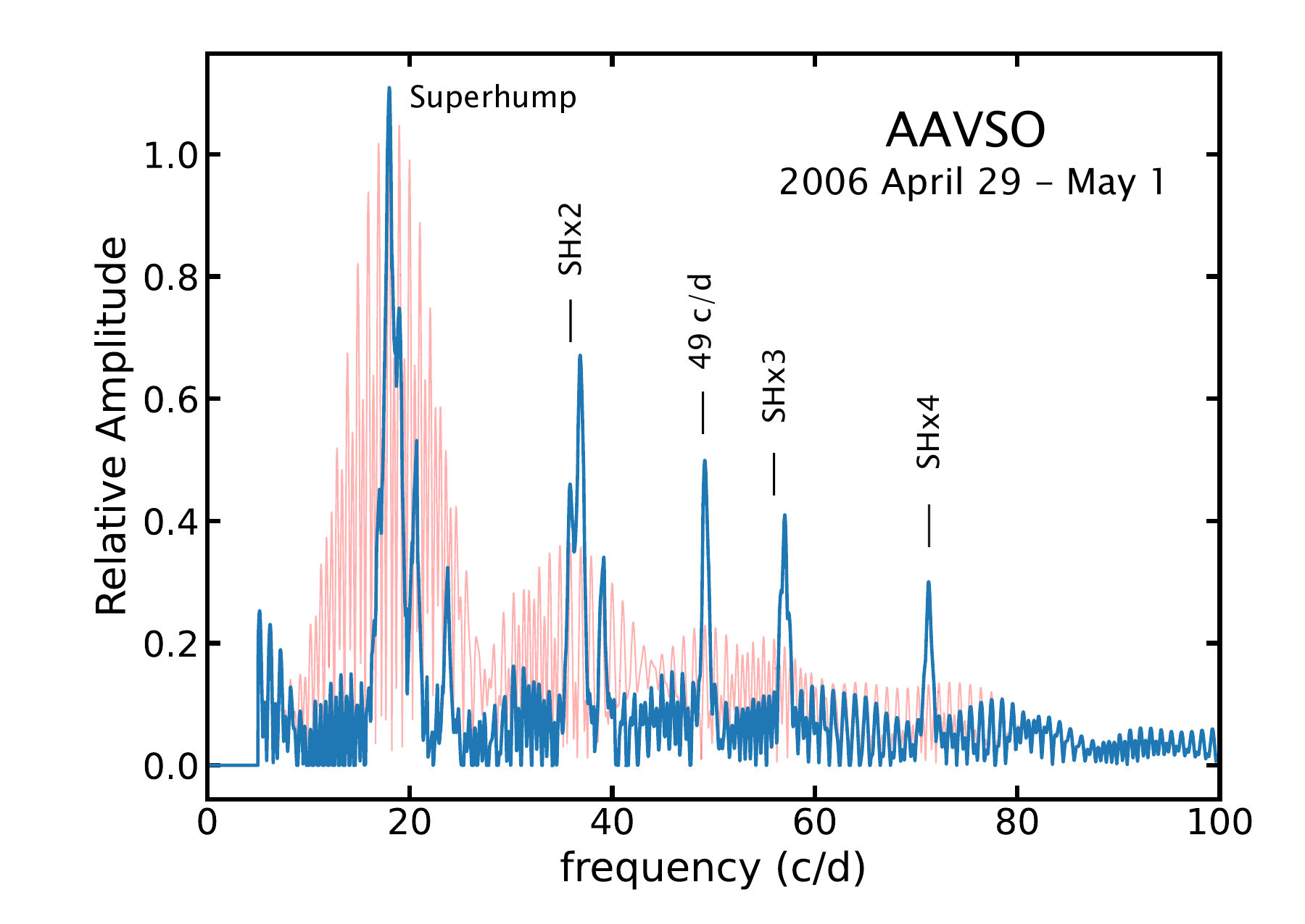}
    \caption{The L-S periodogram of the three nights of V844~Her photometry from the AAVSO database. The faint red line shows the ``dirty'' periodogram resulting from daily gaps in the sampling window. The heavy blue line is the clean spectrum clearly showing the fundamental superhump signal and its harmonics. The 29~min periodicity is recovered in the cleaned spectrum, and its amplitude relative to the SH harmonics is similar to what was seen in the TESS sector~24 data.  }
    \label{aavso}
\end{figure}

\subsubsection{AAVSO}

The 29~min signal was not detected by \citet{thorstensen02} during a 1997 superoutburst despite their analysis of 10 nights of ground-based photometry.  To further test for the presence of the 29~min signal in ground-based data, we analyzed three nights of CCD photometry from the AAVSO database taken five days after the peak of a super-outburst. Despite the high-quality data, the daily gaps in the sampling window resulted in a highly structured L-S periodogram shown in Figure~\ref{aavso}. The fundamental superhump signal and its first harmonic are obvious in the ``dirty'' periodogram. There appears to be a possible signal near 29~min blended with the second superhump harmonic.

To confirm the suspected detections in the dirty periodogram, we applied a cleaning algorithm based on the deconvolution of radio images synthesized from irregularly spaced apertures \citep{clean74}. Here, the dirty beam is generated by calculating the periodogram of the temporal sampling window. The dirty beam is then convolved with the dirty periodogram to identify the frequency of the strongest signal matching the beam shape. The dirty beam is shifted to the frequency of the identified signal, scaled to its peak, and then multiplied by a gain factor before being subtracted from the periodogram. This process is repeatedly applied to the residual periodogram after recording the amplitude and frequency of each subtracted beam. 

The clean periodogram is constructed by placing a scaled Gaussian at the frequency corresponding to each subtracted dirty beam. The residuals remaining after subtraction of all the dirty beams are then added to the synthesized Gaussian periodogram. Adding the residuals is useful in deciding if more clean iterations are required. The results of the cleaning are shown in Figure~\ref{aavso}. The assumed gain was 0.4 and the number of subtracted beams was limited to 12. Because the superhumps vary in frequency with time, the cleaning algorithm identified multiple peaks for the fundamental and major harmonics of the superhumps. 

A 29~min signal is well-detected after the clean algorithm is applied. Given that plausible SH harmonics are detected at lower amplitudes, we conclude that this ground-based data supports the presence of this new periodicity. The amplitude of the 29~min signal relative to the SH harmonics is similar to that seen in the TESS periodogram from early in the outburst (see Figure~\ref{pow}). We note that the 29~min signal was not detected in the extensive photometry obtained during a superoutburst analyzed by \citet{thorstensen02}, yet it was seen in the more modest set of AAVSO data.  Given the difficulty in detecting this signal, future intensive ground-based observations during outbursts are recommended.

\subsubsection{Optical Photometry in Quiescence}

\begin{figure}
    \centering
    \includegraphics[width=\columnwidth]{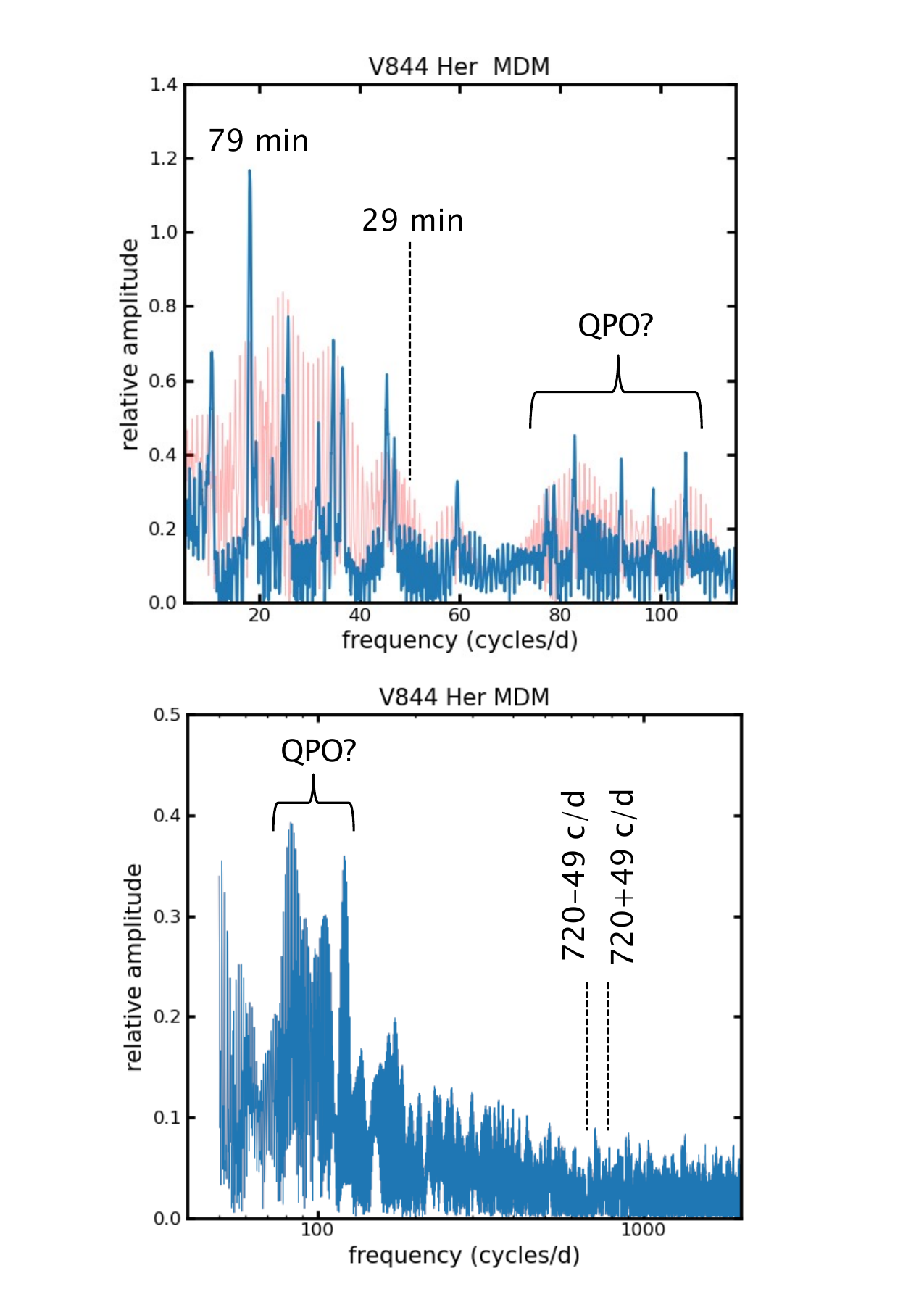}
    \caption{{\bf Top:} The periodogram of four nights of MDM photometry obtained during quiescence. The dirty specturm is shown in light red and the cleaned spectrum in dark blue. No excess power is detected at 29~min. A peak at 79~min likely corresponds to the orbital period of the binary. {\bf Bottom:} A high-frequency extension of the periodogram without cleaning does not show evidence of super-Nyquist frequencies at 671 or 769~\cycleday (1.87 to 2.15~min). Both plots suggest the presence of quasi-periodic oscillations (QPO) corresponding to periodicities between 10 and 15 minutes. }
    \label{mdm}
\end{figure}

We searched for evidence of periodicities in the four nights of MDM data obtained with a 23~s cadence, corresponding to a Nyquist frequency of 1870~\cycleday (46.2~seconds). The L-S periodogram of the full MDM data is shown in Figure~\ref{mdm} and the light curves are reproduced as an appendix. No clear signal associated with the 29~min TESS oscillation is seen in the dirty or cleaned periodograms. Injecting a sinusoid at the 49~\cycleday\ frequency shows that an amplitude of 40~mmag is required to assure detectability. Also, no evidence of a possible super-Nyquist signal is present at high frequencies. Again, by injecting signals with various amplitudes into the MDM light curve, we estimate that a sinusoid with an amplitude of 7~mmag would be detectable with a 3-$\sigma$ significance for frequencies greater than 500~\cycleday (2.88~min). 

The strongest peak in the cleaned periodogram is at 79~min, which is consistent with the orbital period. There appears to be excess power in the frequency range 100 to 150~\cycleday, corresponding to periodicities between 10 and 15~minutes. This suggests the presence of QPOs during quiescence in V844~Her. 

\subsubsection{TESS Sector 51}

TESS began sector~51 in April, 2022, approximately one month after the end of a super-outburst from V844~Her. Based on the ASAS-SN photometry \citep{shappee14, kochanek17}, the super-outburst was caught on the rise March 7, 2022 (HJD=2459646.0) and faded from outburst between March 21 and March 26, 2022 (HJD=2459660 and 2459665). A periodogram using the entire sector reveals no significant signal at 29~min . In sector~24, the 29~min signal had faded below detectability a month after the end of the super-outburst,  so it is not surprising that the 29~min periodicity is not seen in sector~51.

The second harmonic of the orbital period is detected during quiescence in sectors 51 and 52 and might be attributable to the secondary's ellipsoidal variations.

\subsection{X-Ray Properties}

An asynchronously spinning WD in V844~Her could generate a fixed frequency oscillation as this is often seen in IPs.
To check if V844~Her could be an IP, it was compared to two known IPs: V1025~Cen and DW~Cnc. V1025~Cen has an orbital period of 84.6~minutes \citep{buckley98} and DW~Cnc \footnote{An in-depth discussion of DW Cnc's photometric variablity can be found in \citet{ramirez22}.} has a period of 86.1~minutes \citep{patterson04}, both near the period minimum. Both these IPs are also considered low-luminosity IPs \citep{mukai23}, making them likely to be most similar to V844~Her. The X-ray spectra of the three systems were compared using data from \emph{XMM-Newton}.

\begin{figure}
    \centering
    \includegraphics[width=\columnwidth]{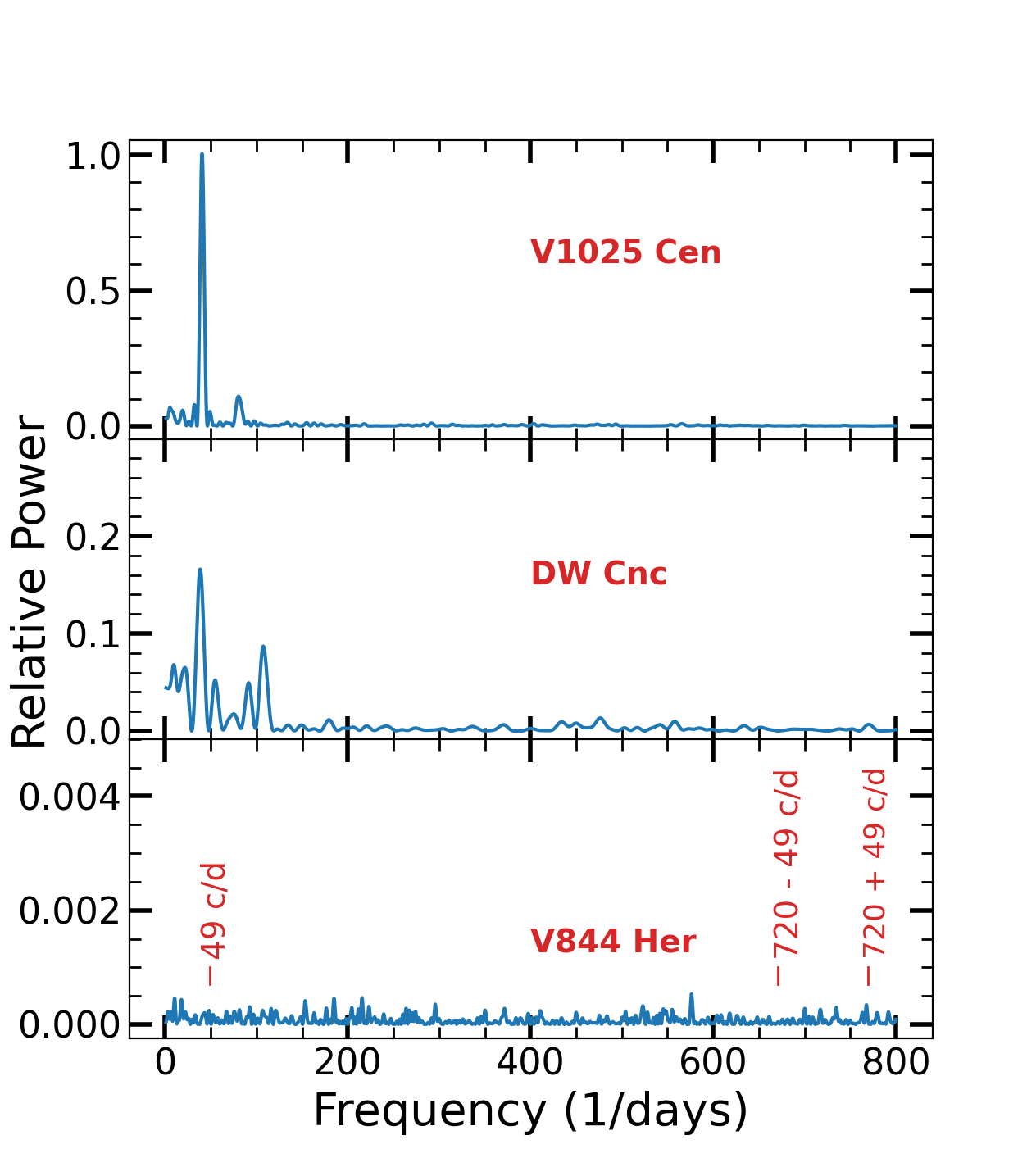}
    \caption{The X-ray periodograms of V1025 Cen (top), DW Cnc (middle), and V844 Her (bottom) using 30-second binning. The spectra are normalized to the strength of the signal in V1025 Cen. A clear peak and subsequent harmonics can be seen in the power spectra of the known IPs, but the spectrum of V844 Her shows no distinctive peak at the 29~min period or possible super-Nyquist frequencies.}
    \label{powerspecIPs}
\end{figure}

\subsubsection{Comparison of X-Ray Light Curves}

The X-ray light curves for V844~Her, V1025~Cen, and DW~Cnc were constructed with 30~s binning of the X-ray count rate. Since the Nyquist frequency is then 1~min, the L-S periodograms were calculated out to that limit, and are displayed in Figure \ref{powerspecIPs}. V1025~Cen shows a strong peak at 40.25~\cycleday , corresponding to the spin period of its WD \citep{buckley98}.  The periodogram of DW~Cnc shows significant peak at its WD spin frequency of 37.3~\cycleday \citep{patterson04}. In the periodogram of V844~Her, no significant peaks are seen at 29~min or at the plausible TESS super-Nyquist frequencies.

\begin{figure}
    \centering
    \includegraphics[width=\columnwidth]{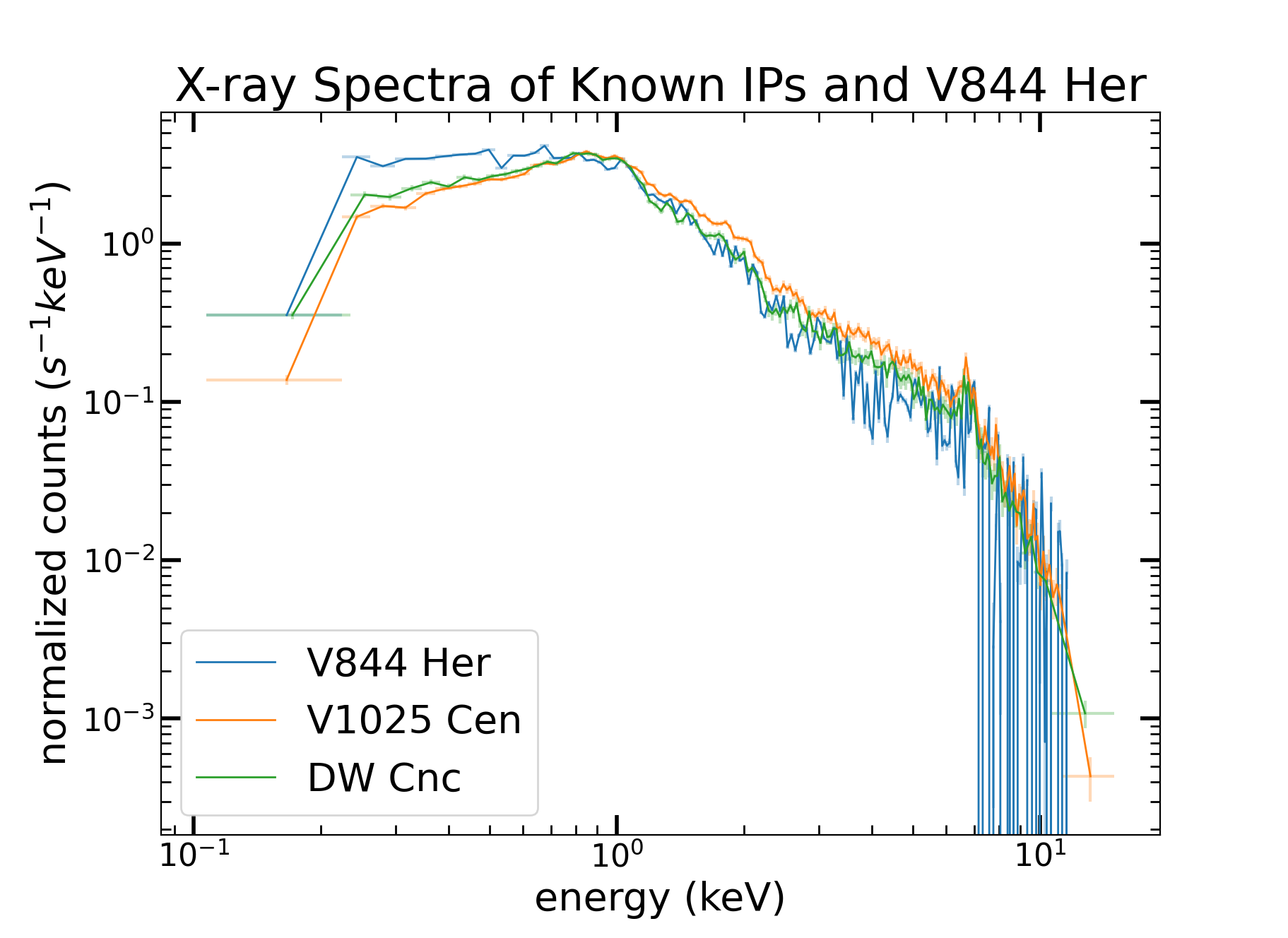}
    \caption{The X-ray spectra of the two known IPs and V844 Her without the bandpass filter applied. The normalized counts~$s^{-1}$~$keV^{-1}$ for V844 Her have been multiplied by 8 to scale the spectrum to the same height as the other two at 1 keV.}
    \label{comparedModels}
\end{figure}

\subsubsection{Comparison of X-Ray Spectra}

Models of the two known IPs were created using three mekal components \citep{mewe85, mewe86, liedahl95} and a Gaussian along with a Tuebingen-Boulder ISM (\emph{tbabs}) absorption model \citep{wilms00}. This allowed just the hydrogen column density to vary. Using XSPEC, the model was created with the command \emph{model tbabs*(mekal+mekal+mekal+gaussian)}.

\begin{deluxetable*}{CCCCCCCCC}
\centering
\tablecaption{Xspec model parameters for the four systems obseved by XMM.  \label{models}}
\tablehead{
\colhead{\text{star${^a}$}} & \colhead{$N_H$} & \multicolumn{3}{c}{mekal Temperature} & \multicolumn{2}{c}{Gaussian$^b$} & \colhead{$\chi^2$} & \colhead{\text{dof}}
\\
\colhead{} & \colhead{} & \colhead{$kT_1$} & \colhead{$kT_2$} & \colhead{$kT_3$} & \colhead{E} & \colhead{$f_g$} & \colhead{} 
\\
\colhead{} & \colhead{($10^{-22}$)} & \colhead{(keV)} & \colhead{(keV)} & \colhead{(keV)} & \colhead{($10^{-4}$)} & \colhead{(keV)}
}
\startdata
\text{V1025 Cen} & 0.050(2) & 10.48(49) & 1.51(15) & 0.671(20) & 0.102(24) & 0.069(63) & 183.6 & 152\\
\text{DW Cnc} & 0.014(2) & 10.87(75) & 1.34(10) & 0.623(24) & 0.163(84) & 0.27(16) & 138.2 & 130\\
\text{V844 Her} & 0.023(7) & 16.7(61) & 1.080(63) & - & - & - & 201.3 & 145 \\
\text{WZ Sge} & 0.014(5) & 11.0(1.1) & 1.28(60) & - & - & - & 112.9 & 109
\enddata
\tablenotetext{a}{The model for WZ Sge was based on data from the MOS1 sensor while all the other models were based on data from the PN sensor. This is because calibration files for the PN sensor during the observation of WZ Sge could not be generated.}
\tablenotetext{b}{For the models that include a Gaussian, the peak was frozen at 6.4 to ensure it represented an iron line.}
\end{deluxetable*}

We used a bandpass of 0.3-10 keV, mirroring the range used in \citet{Kennedy17}. This is due to the energy sensitivity range of the CCD on \emph{XMM-Newton}. For each model, a Gaussian emission line was frozen at 6.4~keV, with the $\sigma$ and norm thawed. To create the model for V844 Her, a two-mekal model with the \emph{tbabs} absorption model was used. Only two mekal components were used because the $\chi^2$ of the fit did not change significantly with the addition of a third. For the known IPs, three mekal components were needed to appropriately fit the soft end of the spectrum below 1.0~keV. A Gaussian was also not used in the model of V844 Her because the spectrum did not show a peak around 6.4~keV. WZ Sge was fit in the same way as V844 Her. The model parameters and normalized $\chi^2$ are shown in Table~\ref{models}.



A comparison of the X-ray spectra is displayed in Figure~\ref{comparedModels}. The three spectra have been shifted to have the same count rate at 1.0~keV. The shape of the spectra are very similar, but V844~Her tends to have a slightly softer spectrum relative to the IPs. 

From the models, the flux was calculated using the XSPEC command \emph{flux errorsims}. The luminosity was then calculated based on GAIA distances and the values can be seen in Table~\ref{photometry}. V844~Her and WZ Sge both have X-ray luminosities much lower than the two confirmed IPs. The luminosity of non-magnetic CVs was found to range between ~$10^{29}$ and $10^{32}$ ergs/s \cite{Pretorius12}, with V844~Her being at the fainter end of that interval.

\begin{deluxetable}{CCCC}
\centering
\tablecaption{The distance, flux, and luminosity of the systems.\label{photometry}}
\tablehead{
\colhead{star} & \colhead{distance${^a}$} & \colhead{flux${^b}$} & \colhead{luminosity}\\
& (\text{pc}) & ($10^{-12}$ \text{ergs/cm}$^2$\text{/s})  & ($10^{30}$ \text{ergs/s})
}
\startdata
\text{V1025 Cen} & 196.0 $^{3.17} _{3.05}$ & 13.1 $^{0.11} _{0.12}$ \>& 60.2$^{2.01} _{1.95}$\>\\
\text{DW Cnc} & 203.1 $^{1.62} _{1.72}$ & 9.81 $^{0.110} _{0.124}$ \>& 48.4$^{0.94} _{1.02}$ \> \\
\text{V844 Her} & 304.9 $^{5.38} _{5.31}$ & 1.01 $^{0.036} _{0.056}$\>& 11.2$^{0.56} _{0.74}$ \> \\
\text{WZ Sge} & 41.18 $^{0.08} _{0.05}$& 11.4 $^{0.107} _{0.151}$ & 2.31$^{0.02} _{0.03}$\>
\enddata
\tablenotetext{a}{Distances from Gaia EDR3 \cite{Bailer-Jones21}.}
\tablenotetext{b}{To calculate flux, the XSPEC command \emph{flux errorsims} was used with a range from 0.3~keV - 10~keV. }
\end{deluxetable}

\begin{figure}
    \centering
    \includegraphics[width=\columnwidth]{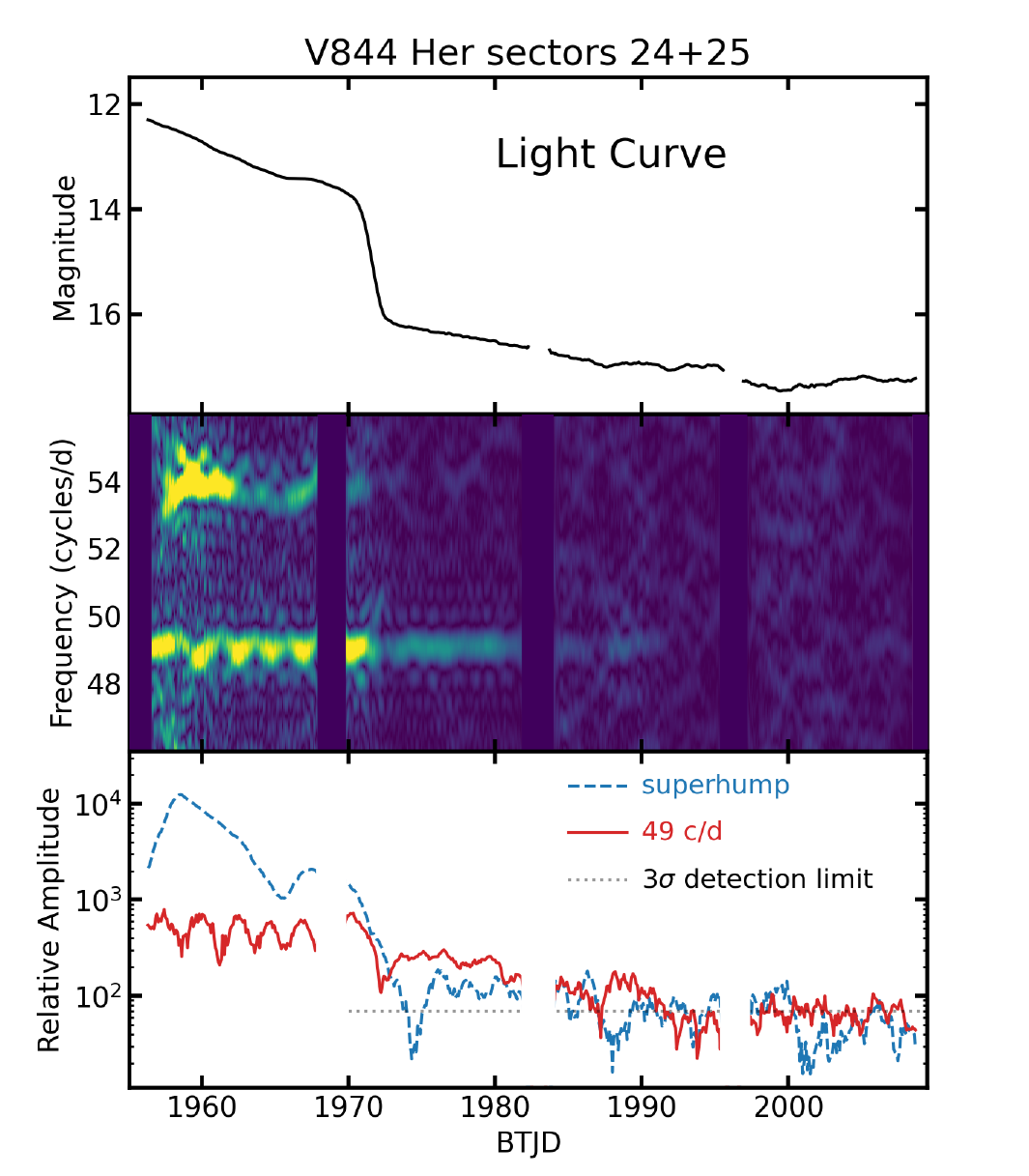}
    \caption{{\bf top:} The light curve of V844~Her over sectors 24 and 25. The average count rate over a 1-day wide moving window has been converted to an approximate magnitude. The outburst ends on BTJD=1972, but the system continues to fade exponentially until BTJD$\approx 1988$. {\bf middle:} The time-resolved power spectrum focused on the frequency region that includes the 29~min signal and the second harmonic of the superhump signal. The 29~min signal shows variability on a 2-day time scale. Data gaps are masked. {\bf bottom:} The amplitudes of the 29~min (solid red line) and 80~min SH (dashed blue line) oscillations from periodograms of a 1-day wide moving window. These oscillations occur at 49 and 18~\cycleday, respectively. Gaps in the TESS data have been masked. The dotted black line indicates a 3$\sigma$ detection threshold for the oscillations. The 29~min oscillation is well-detected out to BTJD$\approx 1991$. }
    \label{amplitude}
\end{figure}

\section{Discussion}

\begin{figure}
    \centering
    \includegraphics[width=\columnwidth]{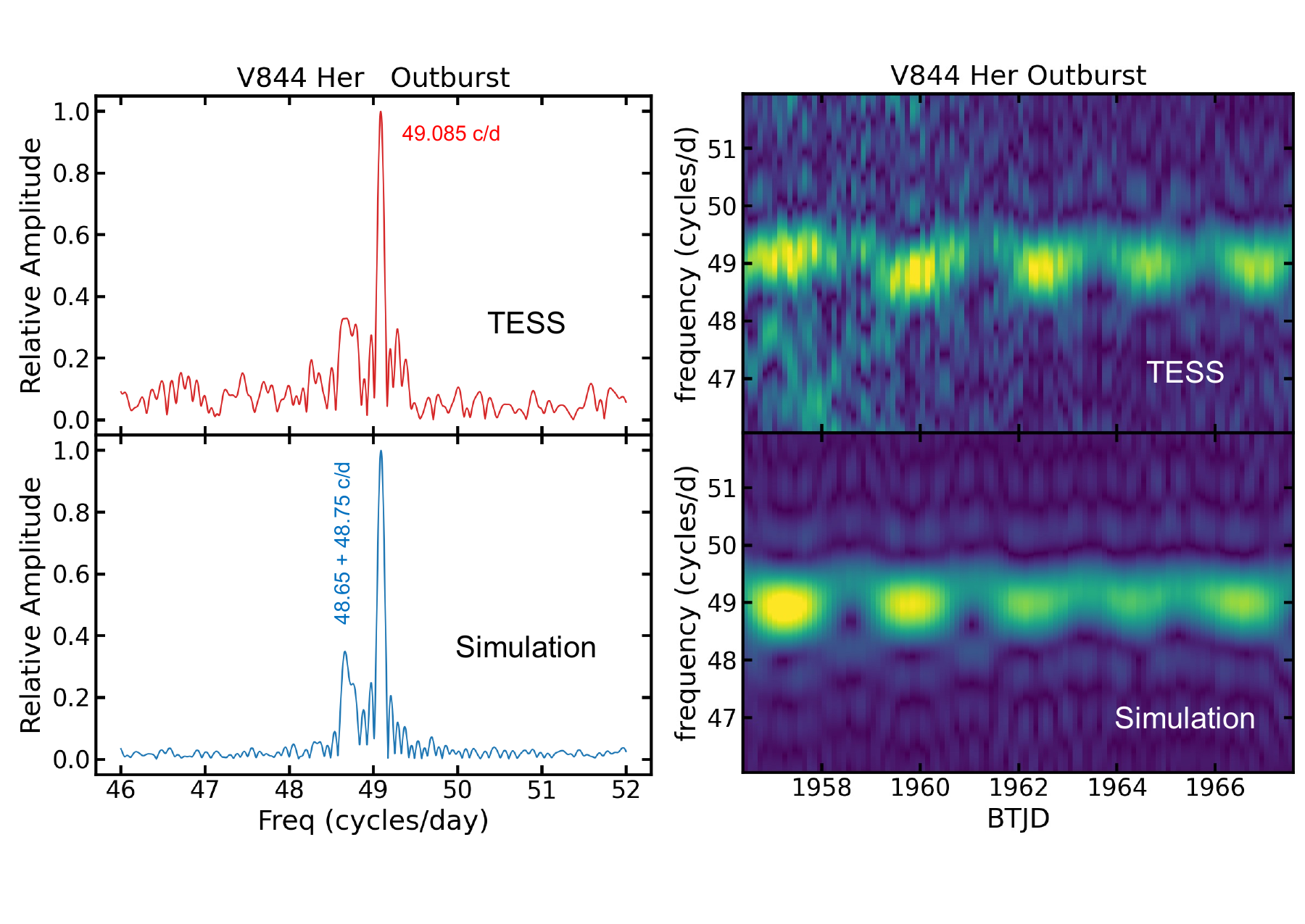}
    \caption{{\bf left:} Periodogram of the TESS data during the outburst (BTJD$<1969$) zoomed in on the 29~min peak (top). The primary signal is concentrated in an unresolved peak at 29.34~mins (49.085~\cycleday). A weak, resolved peak is also present at 29.57~mins (48.7~\cycleday). A periodogram of a simulated light curve is shown at the bottom. The observed beat signal is approximated by adding cosines with frequencies of 48.65 and 48.75~\cycleday\ (corresponding to periods of 29.60 and 29.54~minutes) to a cosine with a frequency of 49.085~\cycleday .  {\bf right:} The time-resolved periodogram of the TESS data (top) during outburst and the simulated light curve (bottom). The simulated light curve shows a modulation similar to that seen in the observations.  }
    \label{simulation}
\end{figure}

\subsection{Behavior of the 29~min Periodicity During Outburst}

Details of the 29~min signal variability are shown in Figure~\ref{amplitude}. The frequency of the 29~min signal is relatively steady when compared with the SH signal. However, there is clearly some variation in its strength during the brightest phases of the outburst. The amplitude of the 29~min periodicity varies on a 2.3$\pm 0.2$-day cycle until the data gap at BTJD=1970. The oscillatory amplitude is suggestive of an interference with a nearby frequency offset from the main oscillation by 0.4~\cycleday . A periodogram of the outburst (BTJD$<1969$) zoomed in on the 29~min signal is displayed in Figure~\ref{simulation}.  The primary peak centered at 29.34$\pm0.01$~min (49.085$\pm 0.005$~\cycleday) is unresolved, that is, its width matches that of a constant frequency sinusoid sampled with TESS time stamps. A weak, resolved signal at 29.54~min (48.75~\cycleday) is also detected and beats against the primary peak to generate the 2.3-day modulation. As shown in Figure~\ref{simulation}, a simulation of the primary signal plus a beat does fairly well at reproducing the observed 0.4~\cycleday\ modulation. 

The origin of a 2.3-day oscillation that could beat against the 29~min signal is not explicitly obvious. One periodicity in the system that does operate on this long time scale is the period of the dynamical SH instability relative to the orbital period.  That is, the difference between the SH period of 80.45~min and the orbital period of 78.7~min, generates a precession frequency of the eccentric disk corresponding to roughly 0.4~\cycleday, or a period of 2.5 days. This could be an explanation for the origin of the oscillation.

SH oscillations are generated primarily in the outer disk, while any impact of a spinning WD or lpDNOs are thought to operate in the inner disk \citep{warner08}. Possibly a spiral instability in the disk during the brightest phase of the superoutburst could connect the outer disk with the region around the WD and generate a beat on the disk precession time-scale \citep[e.g.][]{murray99}.

\subsection{Behavior of the 29~min Periodicity After Outburst}

At the end of the superoutburst (BTJD$\approx 1971$), the amplitude of the 29~min signal drops by a factor of 3 while the SH amplitude fades by a factor of 20 (see Figure~\ref{amplitude}).  After the brightest phase of the outburst ends, the amplitude of the 29~min signal remains nearly constant for about eight days. During this phase, the 29~min oscillation is the strongest periodic signal in the TESS data.

For BTJD$ > 1972$, the flux of V844~Her fades exponentially for approximately 20 days as it returns to quiescence. The 29~min oscillation remains detectable until BTJD$=1991$ in TESS sector~25. 
No clear modulation on the 2.3-day cycle is detected after the outburst fades beyond $\approx 15$~mag.


\subsection{Origin of the New Oscillation: All Things Considered}

We summarize the possible sources of the 29~min signal here:
\begin{itemize}

\item The signal comes from a mechanical or electronic signal originating on the TESS satellite. Our tests show this is unlikely.

\item The signal comes from a star near V844~Her that is contributing light to the extracted aperture. Our tests show this is unlikely.

\item V844~Her is an IP and the signal is the spin period of its magnetic WD. For example, V1025~Cen and DW~Cnc are IPs with short orbital periods and WD spin periods near 30~m.

\item The signal is a Dwarf Nova Oscillation (DNO) or a long-period DNO (lpDNO) \citep{warner04}. DNO typically have periods of around 20~s while lpDNOs are four times longer. These periodicities tend to be shorter than the TESS fast cadence, although they may be detectable through super-Nyquist sampling. 

\item The signal comes from a spiral structure breaking the symmetry of the disk. CVs with small stellar mass ratios  \citep{lin79} or systems with magnetic WDs \citep{murray99} may develop spiral structures in their disks.

\item The signal results from the WD spin or the beat frequency impacting the inner disk or boundary layer. The magnetic field of the WD is likely weak compared to typical IPs, so it is not clear how the WD rotation couples to the accretion.

\end{itemize}

\subsubsection{V844~Her is an Intermediate Polar?}

IPs generally show optical/X-ray oscillations related to the spin period of their magnetized WD. No periodic signal from V844~Her is detected during quiescence at optical or X-ray wavelengths. 
The accretion properties of V844~Her do not appear similar to known IPs with short orbital periods. V844~Her shows classic super-outbursts of SU~UMa type dwarf novae. These outbursts suggest a well-formed Keplarian accretion disk that undergoes episodes of thermal and dynamical instability \citet{kato22}. In contrast, the IPs V1025~Cen and DW~Cnc tend to produce short outbursts ($<$~day) that may be the result of magnetically gated accretion \citep{littlefield22}.

Our analysis of the X-ray luminosity and spectra of V844~Her, V1025~Cen, and DW~Cnc suggest that V844~Her is most similar to SU~UMa type CVs, and unlikely to be a classical IP.

We also compared the X-ray properties of V844~Her with WZ~Sge, the prototype of the sub-class of SU~UMa dwarf novae. WZ~Sge shows rapid oscillations and it has been suspected of being an IP with a very rapidly rotating WD. A similarly rapid signal could appear in the TESS data as a super-Nyquist peak with a period around 30~minutes. However,  we do not detect any high-frequency oscillations in V844~Her that would correspond to the rapid variations seen in WZ~Sge.

\subsubsection{V844~Her Exhibits Dwarf Nova Oscillations?}

DNOs were identified as rapid oscillations with periods ranging from 5-40 seconds \citep{warner08}. They have two time scales on which they vary: one relating to optical brightness and the other to sudden, small changes in period. The periods of DNOs are most consistent at maximum brightness during outbursts, losing coherence as the outburst fades \cite{hildebrand81}. Such lapses in coherence can make signals difficult to detect using Fourier transforms, but can be found using time-resolved power spectra. Less coherent oscillations called QPOs \citep{patterson77} are sometimes associated with DNOs and tend to have frequencies 16 times lower than DNOs.

The 29~min signal is seen during outburst and during post-outburst fading. This is consistent with the  oscillation being a DNO, although, the observed period of about 30~minutes is much longer than DNOs seen in other systems.  The oscillation seen at 29~min might be a high-frequency oscillation seen with the super~Nyquist sampling of TESS. Thus, the true signal originates at 671~\cycleday\ (128.8~seconds) or 769~\cycleday (112.4~seconds). Even these periods near 2~minutes are longer than most observed DNO periodicities, so it may be that we are detecting a long-period DNO (lpDNO). lpDNO tend to have periods roughly four times longer than their associated DNO \citep{warner04}. However, the ground-based detection of the 49~\cycleday\ peak makes it unlikely that TESS sampling is creating a super-Nyquist signal from an oscillation close to the TESS cadence.

We conclude that the source of the 49~\cycleday\ variation is unlikely to be a DNO or a lpDNO.

\subsubsection{V844~Her has a Spiral Disk Structure?}
CVs have been shown to exhibit non-uniform emission due to tidally induced spiral shocks \citep{lin79}. In CVs with high mass ratios, the outer disk is unable to expand to the 2:1 resonance that causes such shocks in low mass ratios. However, the 2:1 resonance can occur within the tidal truncation radius, causing temporary shocks to occur. This is one explanation for the periodicity only being visible during super-outburst.

In IPs specifically, an asymmetric disk structure can alter the accretion on the primary \cite{murray99}. Tidally induced spiral arms propagate through the disk and result in azimuthal asymmetry within the inner disk when the arms extend to the disk-magnetosphere boundary. The accretion rate is boosted as a magnetic pole sweeps past the spiral arm, which can produce power at the spin-orbit beat frequency and related sidebands and harmonics. SU UMa stars develop such a spiral structure during super-outburst, an explanation for the 49~\cycleday\ signal being detectable only during super-outburst.

Unlike SH oscillations, the new signal does not vary in frequency over the superoutburst and it is not clear if a spiral structure would keep its coherence over long periods covered by the TESS data.

The amplitude modulations of the new signal during the first phases of the super-outburst are consistent with the precessional period of the eccentric disk. This suggests a connection between the inner and outer disk that is possible through a spiral wave.

A spot on the inner edge of the accretion disk has been used by \citet{tovmassian07} to account for periodicities seen in FS~Aur and V455~And that appear unrelated to the orbital variations. For V844~Her, such an inner-edge spot would develop only during episodes of enhanced mass transfer. 

\subsubsection{Low B-field Rotating WD?}


\citet{katz75} proposed that a WD magnetic field strength above about 0.1~MG\ would lock the surface field to a crystallized core. This would preclude a slipping equatorial belt of accreted gas from generating DNO phenomena \citep{paczynski78, warner02} and could result in something similar to a low-field IP. \citet{warner08} proposed that WZ~Sge, and its 27.87~s oscillations, is a good candidate for a weak field IP. 

The steadiness of the 29~min signal seen in V844~Her suggests an origin from solid-body rotation. However, the signal is only seen during outburst and as it fades to quiescence. The WD spin signals from IPs are sometimes difficult to see in quiescence. For example, CC~Scl had a 389~s spin pulse discovered during outburst, but it was not visible at optical wavelengths in quiescence. Eventually, the UV light curve from $HST/COS$ did show the spin signal in quiescence due to its higher amplitude at short wavelengths \citep{szkody17}.

\section{Conclusion}

We detect a new photometric periodicity during a superoutburst of V844~Her observed by TESS in sector~24. The variation is not an artifact of the data and its origin is consistent with the position of V844~Her. With a period of 29~min, it is not directly related to the SH oscillations. The X-ray properties of V844~Her are a poor match with IPs with similar orbital periods. The precise cadence of TESS can result in super-Nyquist sampling of rapid oscillations, suggesting that the observed 29~min peak could come from a DNO or lpDNO with periods of 2.15~min or 1.87~min. No detection of a DNO or lpDNO in V844~Her has been previously published. 

The infrequency of superoutbursts in V844~Her, the low-amplitude of the oscillation, and its relatively long period, has made ground-based confirmation of this new oscillation difficult. Our analysis of AAVSO photometry covering three nights during a superoutburst does support the presence of the new signal. Thus, we consider it unlikely that TESS sampling is generating a super-Nyquist signal. However, intensive photometry during future outbursts will be important in determining the nature of the oscillation.

The behavior of the 29~min signal is fairly complex over the super-outburst and subsequent fading. While the SH oscillations vary in amplitude by more than a factor of 10 over the brightest phases of the outburst, the amplitude of the 29~min signal is relatively steady in comparison.  During the initial two weeks of the outburst, the amplitude of the 29~min signal modulates over a 2.3-day cycle that appears to be a beat frequency offset by 0.4~\cycleday\ from the 29.34~min periodicity. The precession cycle of the eccentric disk relative to the secondary star operates on an approximately 2.5~d period, but it is not clear how the disk precession is physically connected with the 29~min variation.

During the final fading to quiescence, the 29~min becomes the dominant signal in the TESS light curve.

Given the steady frequency of the 29~min signal in the TESS data, we suggest that the oscillation is most likely caused by the rotation of the WD in V844~Her. A weak magnetic field on the WD may generate a photometric modulation during outburst but it may be difficult to detect during periods of disk quiescence. That is, V844~Her falls short of being an IP, but the asynchronous rotation of the mildly magnetic WD may generate a spin or beat pulse detectable through outburst and fading to quiescence. Deep photometry during quiescence at blue or ultraviolet wavelengths would provide a test of this hypothesis.

\bigskip

\begin{acknowledgements}

We acknowledge with thanks the variable star observations from the AAVSO International Database contributed by observers worldwide and used in this research. Some of the data presented in this paper were obtained from the Mikulski Archive for Space Telescopes (MAST) at the Space Telescope Science Institute. The specific observations analyzed can be accessed via \dataset[DOI]{http://dx.doi.org/10.17909/yc7m-vr65}.

We acknowledge that this research was funded by a College of Science Summer Undergraduate Research Fellowship (COS-SURF) from the University of Notre Dame. This research was partly funded through NASA grant 80NSSC22K0183.

\facilities{AAVSO, TESS, McGraw-Hill (Templeton), XMM}

\end{acknowledgements}

\begin{appendix}

\begin{figure}
    \centering
    \includegraphics[scale=0.9]{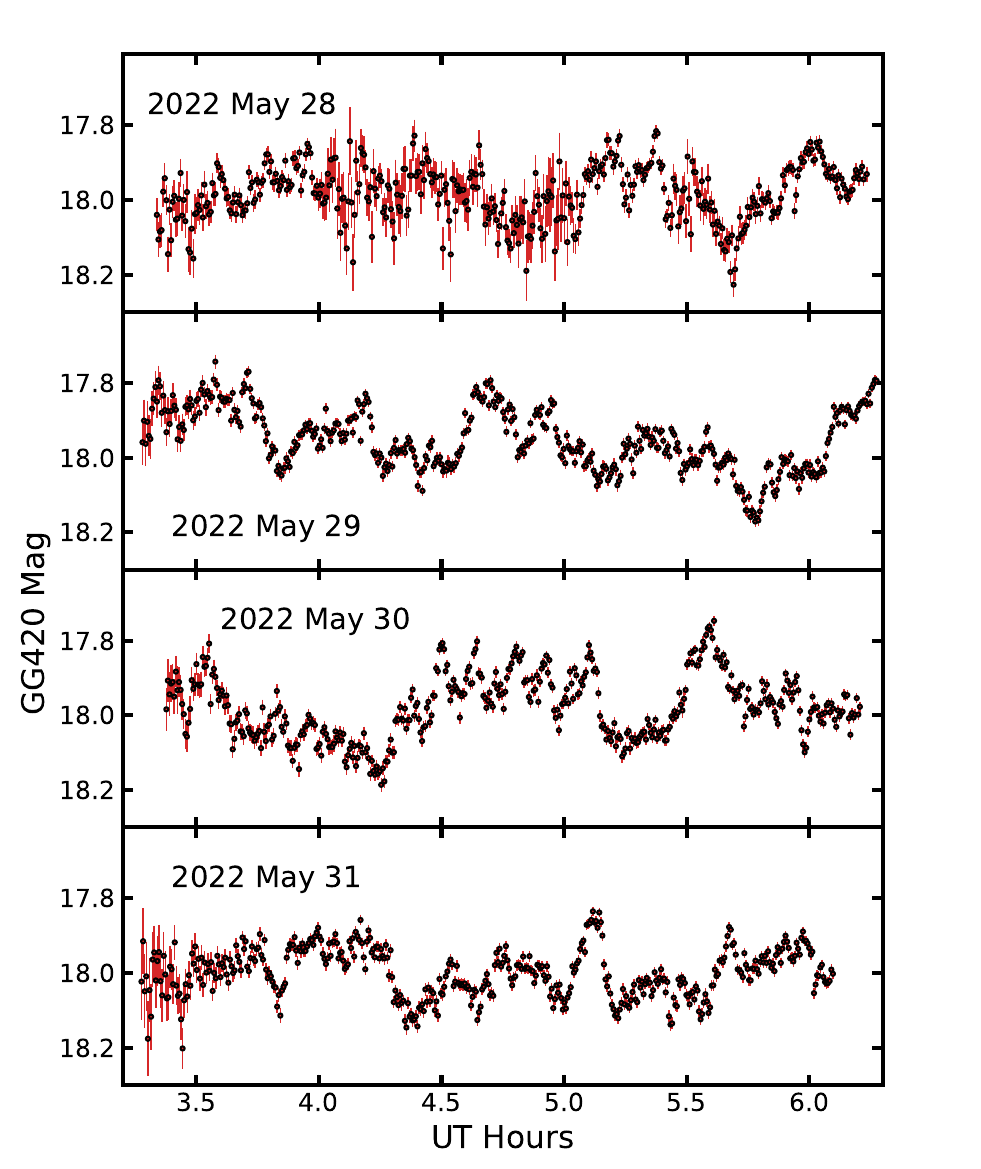}
    \caption{The MDM lightcurves of V844~Her from the four individual nights. The cadence of the time series is 23s. The system is seen to vary in quiescence with an amplitude of $\pm 0.2$ mag.}
    \label{mdmLightCurve}
\end{figure}

The full light curves of V844~Her obtained at MDM over four nights are displayed in Figure~\ref{mdmLightCurve}.

\end{appendix}




\end{document}